# Developing Engineering Model Cobra fiber positioners for the Subaru Telescope's Prime Focus Spectrometer

Charles Fisher*[a], Chaz Morantz[a], David Braun[a], Michael Seiffert[a], Hrand Aghazarian[a], Eamon Partos[a], Matthew King[a], Larry Hovland[a], Mark Schwochert[a], Joel Kaluzny[a], Christopher Capocasale[a], Andrew Houck[a], Johannes Gross[a], Dan Reiley[b], Peter Mao[b], Reed Riddle[b], Khanh Bui[b], David Henderson[c], Todd Haran[c], Rob Culhane[c], Daniele Piazza[c], Eric Walkama[c]

[a]Jet Propulsion Laboratory/California Institute of Technology, 4800 Oak Grove Dr, Pasadena, CA, 91109 USA; [b]Caltech Optical Observatories, 1201 E. California Blvd, Pasadena, CA 91125 USA; [c]New Scale Technologies, Inc, 121 Victor Heights Parkway, Victor, NY 14564 USA

## ABSTRACT

The Cobra fiber positioner is being developed by the California Institute of Technology (CIT) and the Jet Propulsion Laboratory (JPL) for the Prime Focus Spectrograph (PFS) instrument that will be installed at the Subaru Telescope on Mauna Kea, Hawaii. PFS is a fiber fed multi-object spectrometer that uses an array of Cobra fiber positioners to rapidly reconfigure 2394 optical fibers at the prime focus of the Subaru Telescope that are capable of positioning a fiber to within 5µm of a specified target location. A single Cobra fiber positioner measures 7.7mm in diameter and is 115mm tall. The Cobra fiber positioner uses two piezo-electric rotary motors to move a fiber optic anywhere in a 9.5mm diameter patrol area. In preparation for full-scale production of 2550 Cobra positioners an Engineering Model (EM) version was developed, built and tested to validate the design, reduce manufacturing costs, and improve system reliability. The EM leveraged the previously developed prototype versions of the Cobra fiber positioner. The requirements, design, assembly techniques, development testing, design qualification and performance evaluation of EM Cobra fiber positioners are described here. Also discussed is the use of the EM build and test campaign to validate the plans for full-scale production of 2550 Cobra fiber positioners scheduled to begin in late-2014.

## 1. INTRODUCTION

The Prime Focus Spectrograph (PFS) is a fiber fed multi-object spectrometer being developed for the Subaru Telescope on Mauna Kea, Hawaii[1]. PFS forms half of the SuMIRe (Subaru Measurement of Images and Redshifts) project along with the Hyper-Suprime Cam. PFS uses an array of Cobra fiber positioners to reconfigure 2394 optical fibers to within a 1σ accuracy of 7 microns of an astronomical target in less than 105 seconds.

The Cobra fiber positioner[2] uses a θ-φ configuration to place an optical fiber anywhere within a 9.5mm diameter patrol circle. Both Stage 1 (θ) and Stage 2 (φ) motors are driven by ultrasonic rotary piezo tube motors that were developed by New Scale Technologies, Inc.[3] Specifications for the Cobra fiber positioner are listed in Table 1. The Cobra fiber positioner has been in development since 2008 with various iterations of prototypes aimed at improving performance and meeting PFS system requirements. The development history is highlighted in Table 2. Since the original concept the Cobra positioner has seen improvements in reduced voltage for driving motors, more uniform performance with material selection, and tighter alignment of the optical fiber to the focal plane.

*charles.d.fisher@jpl.nasa.gov; phone 1 818 393-5067; http://www.jpl.nasa.gov/

Table 1 - Cobra design specifications

| | | |
|---:|---:|---|
| Outer Dia. | 7.7 | mm |
| Inner Dia. | 1 | mm |
| Length | 80 | mm |
| Motor Offset | 2.375 | mm |
| Patrol Dia. | 9.5 | mm |
| Stage 1 Range of Motion | 370 +/- 5 | deg |
| Stage 2 Range of Motion | 180 +0/-5 | deg |
| Min. Stall Torque | 300 | uNm |
| Stage 1 Step Size | 0.06 | deg |
| Stage 2 Step Size | 0.12 | deg |

Table 2 - Cobra development history

| | Gen0 | Gen1 | Gen2 | Gen3 | EM1 | EM2 | Production |
|---|---|---|---|---|---|---|---|
| **Stage 1 motor size** | 4.4mm | 4.4mm | 3.4mm | 3.4mm | 3.4mm | 3.4mm | 3.4mm |
| **Stage 2 motor size** | 2.4mm | 2.4mm | 2.4mm | 2.4mm | 2.4mm | 2.4mm | 2.4mm |
| **Motor voltage** | 100V | 100V | 175V | 8V | 10V | 10V | 10V |
| **Tilt Req't** | n/a | n/a | ≤ 0.3° | ≤ 0.3° | ≤ 0.3° | ≤ 0.3° | ≤ 0.3° |
| **ΔFocus Req't** | n/a | n/a | ≤ 25µm | ≤ 25µm | ≤ 25µm | ≤ 25µm | ≤ 25µm |
| **Quantity** | 2 | 5 | 2 | 2 | 19 | 11 | 2550 |
| **Delivery Date** | Dec 2008 | May 2012 | Jun 2013 | Aug 2013 | Jan 2014 | March 2014 | *late-2014* |

To facilitate the construction of PFS the Cobra has been designed to be assembled into 42 identical modules that each contains 57 Cobra fiber positioners[4].

## 2. ENGINEERING MODEL

Prior to entering full-scale production of the Cobra fiber positioner an Engineering Model (EM) development phase was undertaken to validate the Cobra design specifications and assembly techniques using processes consistent with large scale production at both New Scale Technologies and at JPL/CIT. The EM also served as a dry run to identify difficult portions of the assembly and test flow and allow for more efficient or effective improvements to the processes. One example of a problem uncovered during the EM effort was solder joint failures between flex print cables and the piezo plates on the motors. To correct the failure mode, solder joints were replaced with an electrically conductive film adhesive. The technique was shown to be more reliable thru extensive qualification of the procedures and tooling design. Qualification testing included temperature and humidity exposure prior to performance testing of the motors.

During the prototype developments some of the motors became non-functional after lubrication oil from the bearings migrated into the friction interfaces of the motor. Using non-lubricated bearings was base-lined instead of investing an extensive amount of time developing an application technique that would put a very small amount of lubricant onto the bearings that would still have a non-zero likelihood of still migrating to undesirable areas. The risks of not using lubrication in the bearings were increasing the parasitic drag in the system, which would lower torque margins, and the potential to shorten lifetime due to corrosion or wear in the bearings. To test the corrosion and wear concerns 5 each of Stage 1 and Stage 2 units were assembled without piezo motors, but all other components (bearings, housings, shafts, etc). They were exposed to +60C and 95% humidity in ambient lab air (no additional corrosive agents added) for a total of 8 days. This exposure is the equivalent of 10 years of exposure at the summit of Mauna Kea using equivalent corrosion damage calculated using ISO9225 guidelines. Average Mauna Kea weather was assumed to be +4C and 11% RH, which was based upon a one year sample of weather station data reported by the Subaru Observatory. After the 8 days of high humidity exposure the bearings were cycled for 500k revolutions in both directions. All but two test units showed negligible increases in drag torque at the end of testing compared to values measured prior to any exposure. The two units that did have increased drag torque were disassembled and found to have no foreign contamination based upon chemical analysis. It is possible that the results were false negatives since galvanic corrosion between the bearings and other components was possible due to the elevated temperature and humidity that would not be possible in the much colder, dryer environment of Mauna Kea. The root cause of increased torque was never determined, but proceeding without lubrication was determined to be low risk.

As the EM design was being prepared for fabrication and assembly it became clear that the unit cost for full scale production would be too great for PFS to afford. It was then decided to split the EM into two phases: EM1 and EM2. The EM1 phase would provide baseline performance metrics for a design that was similar to the earlier prototypes. The EM2 phase would focus on cost-reduction and final system verification.

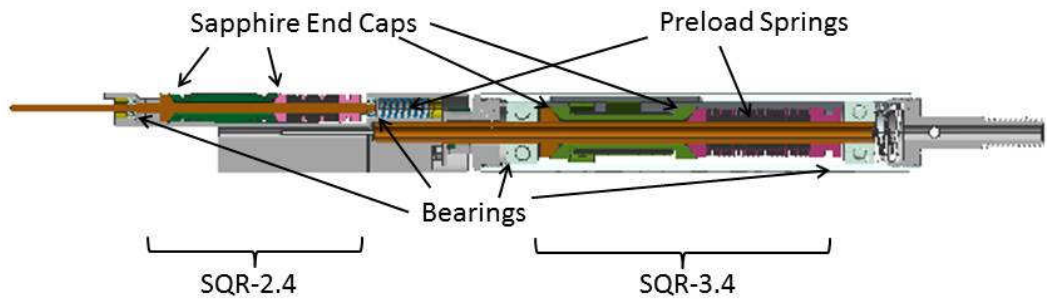

Figure 1 - EM2 Cobra Configuration

### 3. EM2 DESIGN

The most costly items in the EM1 design were the alumina ceramic end caps of the motors, the 2mm OD bearings used in Stage 2 and the motor pre-load bellows spring, which are shown in Figure 1. To reduce the unit price the pre-load bellows were replaced by machined helical springs and the 2mm OD bearing were replaced by 2.5mm OD bearings. There were two promising alternatives to the alumina ceramic: sapphire and nitride steel. A series of down-select tests were performed on motors built using alumina, sapphire and nitride steel. Motors of each type were cycled 100k revolutions in each direction at room temperature and tested for resonant frequency and stall torque as well as inspected for wear. All three materials met design specs and the nitride steel provided the best performance in terms of stall torque, but generated substantially more debris than the alumina or sapphire. The sapphire was the lower cost option compared to alumina so sapphire was base-lined for the EM2 build.

A variety of reliability issues were discovered from the EM1 and EM2 testing that needed to be addressed prior to a final life test and the production build. The two most significant issues were end caps de-bonding from motors and flex print cables rubbing and shorting out. The former was solved by changing adhesives and bonding preparation. The later was solved by adding an extra layer of polyimide to the portions of the cables prone to rubbing and taking greater care in forming and routing them during assembly.

### 4. LIFE TEST

Two EM2 Cobra units were assembled and used for life testing. Their resonant frequencies stall torques and step size performances were characterized prior to any cycling. The units were then cycled to 300k cycles at -5C with intermediate performance tests conducted at the 100k and 200k cycle milestones. For the purpose of this testing a cycle is defined as a clockwise rotation from a hard stop to a random angle within the range of motion and a counter-clockwise rotation back into the hard stop with a 1s stall into the hard stop. A lifetime for PFS is 250k cycles, which includes 10 years' worth of science configurations (~65k cycles) and significant ground testing and day time calibration (185k cycles).

Around 265k cycles into testing one of the test units had both motors cease to run. At that time it was noticed that a motor control parameter for the Stage 2 motors had been incorrectly set after re-starting the tests at 200k cycles. This caused the stage 2 motors to stall into their hard stops for 5 seconds per cycle instead of 1 second. In terms of motor run time, this was the equivalent of putting an additional 260k cycles on the motors. Both tests units were removed for further inspection. On one unit the Stage 1 motor did not show a resonant frequency and upon further inspection had cracking that generated large debris from the sapphire end caps, which is shown in Figure 2. The end caps were also partially de-bonded, which explained the loss of resonance. The Stage 2 motor still performed above design specification, but stall torque had dropped by half. Since a total of >500k equivalent cycles were applied to the motor, the motor was considered passing for the lifetime requirement. This unit was removed from the life test due to the Stage 1 failure. The other unit showed no signs of damage or degraded performance so it continued on with testing to 300k cycles without any issues.

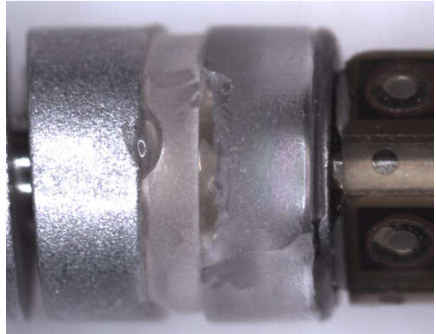

Figure 2 - Damaged end cap from life test

The Stage 1 sapphire cracking was a significant failure mode that had never been observed prior to the testing. A non-housed 3.4mm motor was probed with a laser vibrometer while being run with the same motor settings from the life test (0.5ms "burst" to the motor every 5ms). Significant axial motion of the motor pre-load spring was observed at a frequency of ~400Hz with amplitudes >12µm. A displacement of 12µm at the motor end cap would result in gapping, which causes a jack-hammer effect of the two sapphire pieces impacting each other. Mechanical analysis of the pre-load spring confirmed a natural frequency in the axial direction of ~450Hz. The laser vibrometer data showed that the springs axial resonance was excited by the 65kHz drive frequency of the motor. A test was conducted to see if there were motor control parameters that would get adequate motor control without exciting the end cap gapping mode. The planned durations of motor bursts to be used for PFS are between 0.04ms and 0.3ms, which are shorter than the 0.5ms bursts used for the life testing. The length of the burst given to the motor was varied from 1ms down to 0.1ms with the displacement of the end cap on the preload spring measured by the laser vibrometer. Results from testing with 1ms and 0.25ms bursts are shown in Figure 3. The blue trace is the motor drive signal, the pink trace is the current draw of the system and the yellow trace is the laser vibrometer measurement of amplitude. All signals have a 5kHz bandwidth filter applied to eliminate noise from the 65kHz drive frequency. The amplitude in the yellow trace of the 1ms test is ~25µm, which demonstrates unacceptable jack-hammering. The 0.25ms test had no measureable amplitude. All tests below 0.33ms were consistent with the results shown in the 0.25ms test. Since the PFS use case will not drive motors with bursts larger than 0.3ms this failure mode is likely to be mitigated. A new life test during the early phase of full scale production with PFS specific drive values is expected to demonstrate longer life time where this jack-hammer failure mode is non-existent.

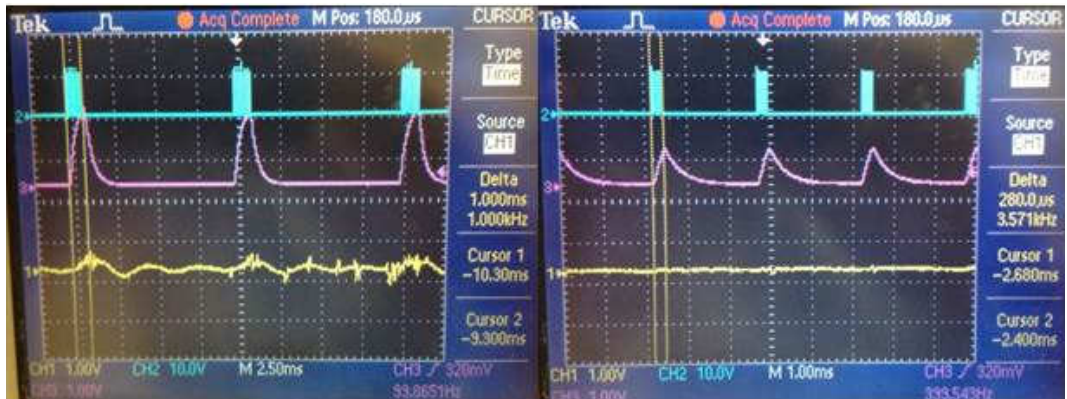

Figure 3 - Motor end cap response due to 1ms (left) and 0.25ms (right) drive signal bursts; blue = motor drive signal, pink = current draw, yellow = laser vibrometer measurement of amplitude

## 5. EM MODULE ASSEMBLY

A total of 19 EM1 units and 11 EM2 units were produced. All 19 of the EM1 units and 9 of the EM2 units were delivered to JPL for assembly into a module and further performance testing. The two that were not delivered were used for life testing. The receiving and assembly process at JPL was intended to mimic the workflow expected for production. The basic steps and work flow are shown in Figure 4.

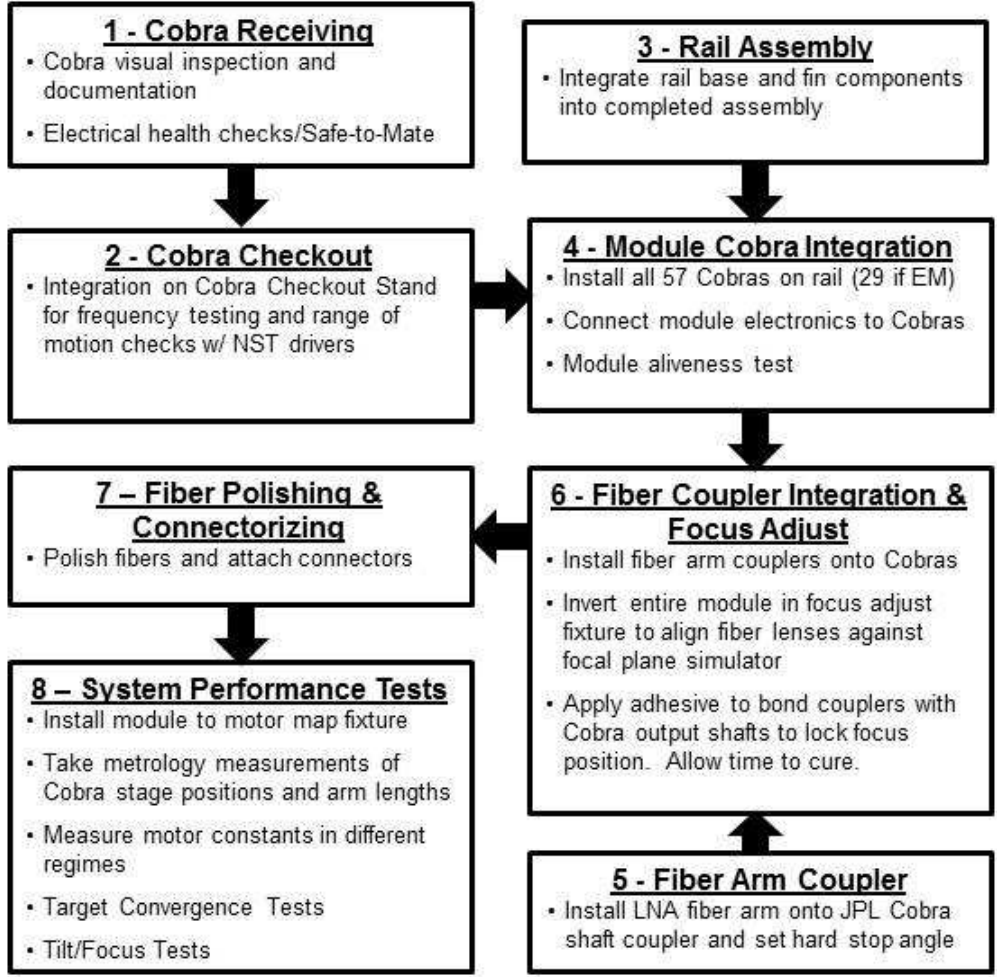

Figure 4 - EM module assembly workflow

The most critical step in the work flow was setting the focus position of all fibers on the module to the same level. The difficulty in this task is that the overall length of the Cobra cannot be controlled to the <10µm level necessary for PFS without a great deal of cost and complexity. To compensate for the variations in the Cobra length, fiber arm height and module thickness a fixture was designed that had the overall length control features machined into the same piece. Precision machining allowed the fixture to be made to a tolerance of 5µm for planarity and parallelism. An optical flat with flatness <1µm was used as a focal plane surrogate. The piece that corrects for the variations in Cobras is the Shaft Coupler, which is free to slide along the length of the Stage 2 output shaft before being bonded in place onto the output shaft. All of the shaft couplers, with fibers already installed, are placed on the Cobras followed by the optical flat being installed onto the fixture. The entire assembly is then flipped over such that gravity causes the microlens of each fiber arm to rest on the optical flat. The shaft couplers are then bonded to the output shafts in this orientation. After setting the focus position of the fibers, the module is then installed into a test stand for system performance tests: fiber tilt and focus errors, and target convergence performance.

# 6. SYSTEM PERFORMANCE TESTING

To accurately characterize the performance of the EM Cobra positioners, a testbed was constructed at JPL with a separate testbed constructed at CalTech. The JPL testbed was configured to test 5 Cobras at a time. The Caltech testbed was designed to test an entire module. The EM module assembly is mounted on a fixture that also contains several fiducial fibers. The fiducial fibers and positioner fibers (science fibers) are then imaged by a camera to ascertain metrology data and simulate science target convergence. The JPL setup is shown in Figure 5.

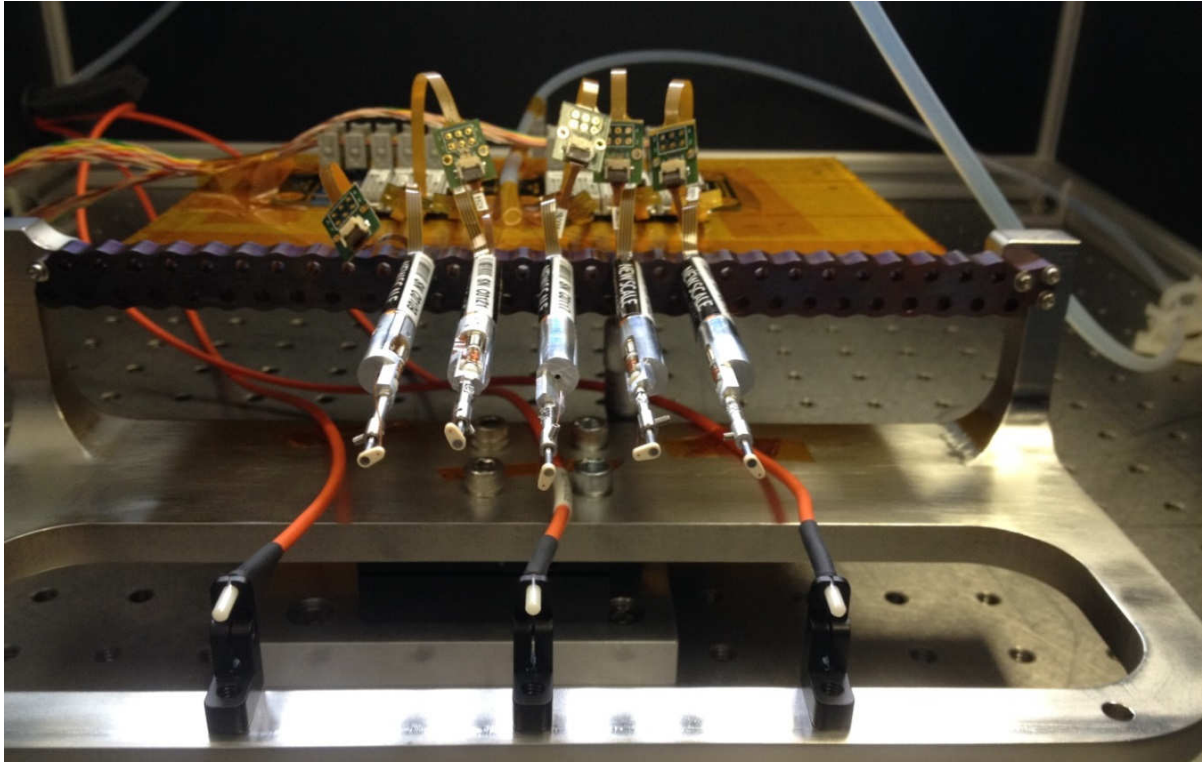

Figure 5 - JPL EM testbed configuration

The first step in calibrating this setup was to establish references for a universal coordinate frame in which the data would be gathered for each positioner. This was accomplished via three fiducial fibers, seen in orange jacketing in Figure 5. The fiducials allow the rotation, translation and scaling to be determined, which are necessary to transform each image's centroids from CCD coordinates to a consistent set of coordinates. Horn's[5] method for absolute orientation was used to calculate the transformations. Fiducial centroids were repeatable to <0.25µm (1s), much less than the target convergence goal of 5µm.

Five geometric parameters needed to be measured for each positioner before target convergence testing could begin: 1) Stage 1 rotation centers, 2) the offset between Stage 1 and Stage 2 rotation centers, 3) the offset between the science fiber and Stage 2 rotation center, 4) the hard stop to hard stop range of motion for each motor, and 5) the orientation of the Stage 1 hard stops. These parameters were routinely gathered during the initial EM installation and also during daily testing. Rotation centers and offsets were determined by analyzing long exposures during which the back illuminated fiber was strobed between small moves of each motor. This resulted in an image of dots in a circular pattern as seen on the left side of Figure 7. A circle was fit to each set of dots by least-squares. The resulting centers and radii are used to calculate the geometric parameters for each motor. The positioner orientation and ranges of motion come from separate images of back illuminated fibers while the motors are at their forward or reverse limits.

## 6.1. Motor characterization

When the positioners arrive from the manufacturer, the burst times for each direction need to be calibrated for each motor to ensure the motor can cover the full range of travel and also make small enough steps to converge to a target. An automated method was used to do this concurrently for all motors on the module. The motors are run with increasing burst times. For each burst time, an image of the resulting motor motion is captured by taking a long exposure that spans the full time the motor is in motion. The first burst time at which a positioner makes a full revolution in 6000 steps is used as the burst time for all future testing. An example of burst time testing is shown in Figure 6. The motor in the center makes a full revolution, while the other four motors only make partial revolutions.

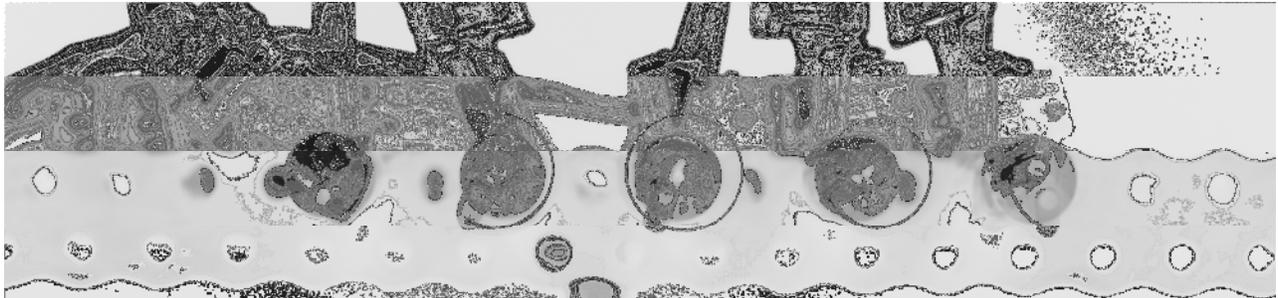
Figure 6 - Example motor streak image for burst time testing

Understanding the variations in angle travelled with a given burst time is key to rapid target convergence. One method used takes a long exposure while commanding a motor to move a fixed number of steps. After each series of steps, while the motor is stopped, the back illumination LED is pulsed to records its position on the image. These moves and LED pulses are repeated over the full range of motion of the motor. On the left-hand side of Figure 7, the result of a pulsed LED experiment is shown. It was possible to create a table of angular displacement versus angular position by calculating the angles between adjacent fiber positions.

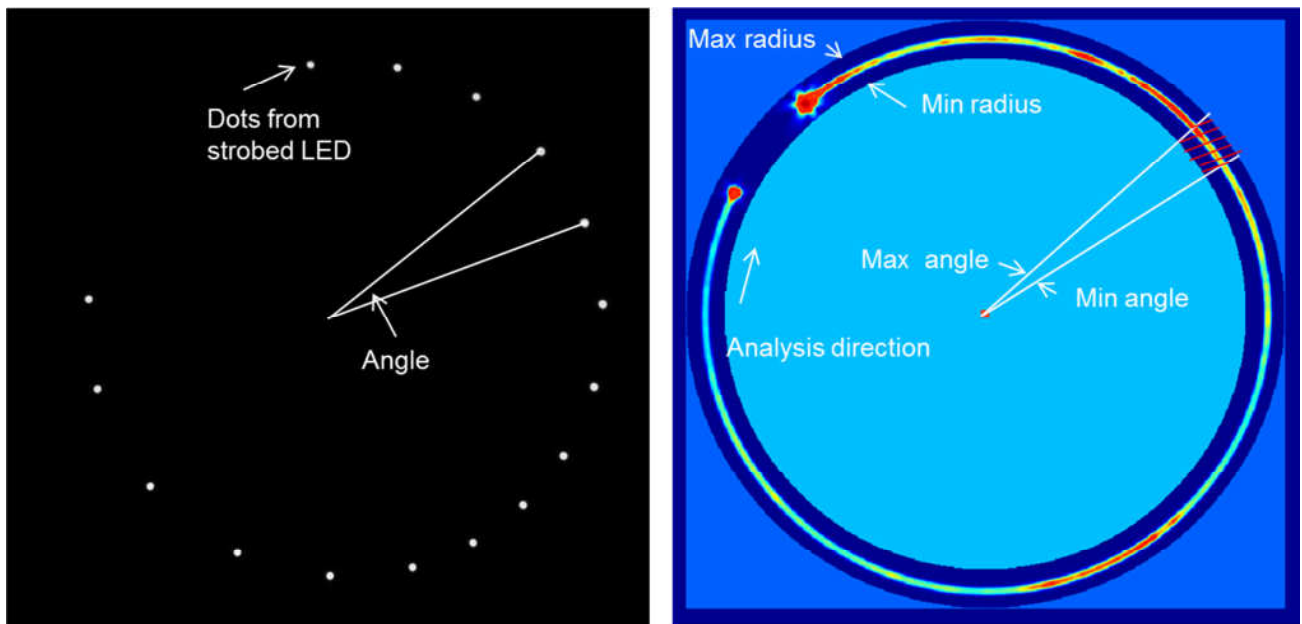
Figure 7 - Example results from LED pulses (left) and streaks (right)

A method was developed where the LED stayed on over the full range of motion of the motor to measure the changes in angular velocity with higher resolution. The right side of Figure 7 shows a processed image of the streak from such a test. The image shows the inner and outer bounds (min and max radius) of where the light is measured. The streak is divided into angular bins (e.g. between the min and max angle) and the counts of all pixels within one bin are summed up. Pixels with a higher intensity (shown in red) indicate a lower velocity. The velocity is calculated by:

$$v = \frac{bf}{c} \qquad (1)$$

where v is the velocity {degrees/step}, c is the integrated counts per bin, b is the bin size {degrees}, and f is the flow rate {counts/step}. From this equation, it can be seen that this method introduces other dependencies on measurements, such as the flow rate of photons hitting the camera detector. This flow rate is the actual number of photons that reaches the camera from one fiber per time (length of a step). It is measured by integrating the signal (counts) from short (0.05s) exposures. The method is sensitive to background lighting. The higher resolution and fast data generation of the streak method resulted in many insights on motor performance. By increasing the burst times for a given motor over several measurements, it was observed that a higher burst time can level out peaks in the velocity curve. This means the motor performs more consistently with a higher burst time, but the angular resolution is reduced. From generating many of the streak images (e.g. 700 over a weekend) motor performance over time could be tracked, which offered insights into motor behavior. Some motors change their velocity profile if they are used too frequently, possibly due to thermal changes. Over a longer period of time, the streak data demonstrated that certain positioners exhibit signs of wear in in some parts of the motor travel, changing the shape of the velocity versus angle curve.

Figure 8 shows the resulting velocity versus angle curves from a set of streaks. It should be noted that the velocity for forward and reverse movements are generally not the same.

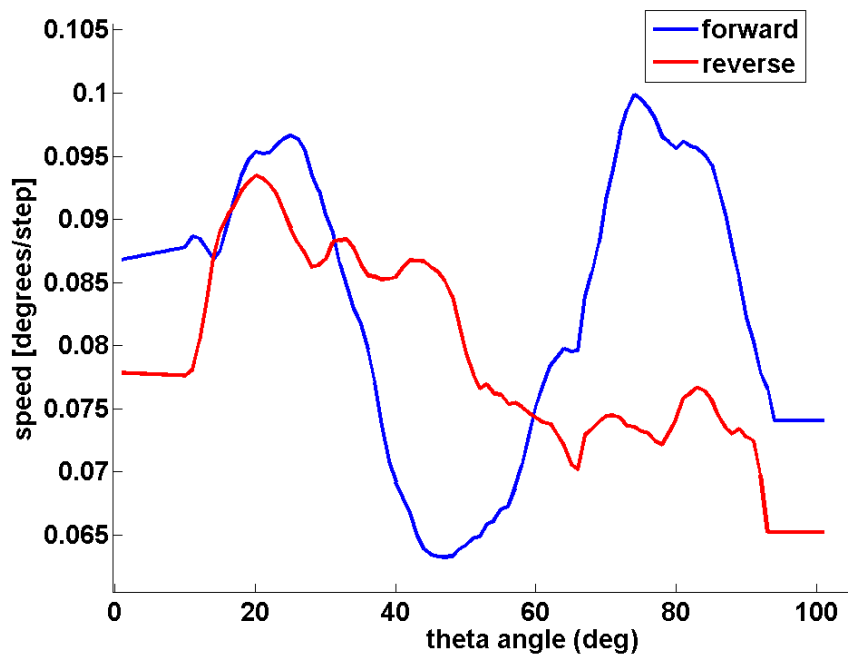

**Figure 8 - Example motor velocity curves**

### 6.2. Target convergence testing

To simulate science operations of the cobra positioners, 100 random targets were selected within the patrol region of each positioner. Each positioner is allowed a maximum of 10 iterations to arrive within 5µm of the target. An iteration consists of: 1) taking an image, 2) determining centroids of the fibers, 3) computing motor angles based upon the centroids and the geometry parameters determined earlier, 4) determine the distance needed to travel to the target, 5) calculate the number of motor steps needed to move to the target based upon the motor velocity curves, 6) command the motors to move, and 7) repeat steps 1-6 until fiber is within 5µm of the target position.

The results from each target test were analyzed several ways. The first look involves a simple plot, seen in Figure 9, of all targets in the specific positioner's patrol region with color representing the number of iterations it took to converge. This plot allows for a quick visual check to see if a positioner has problems converging onto a target in certain areas.

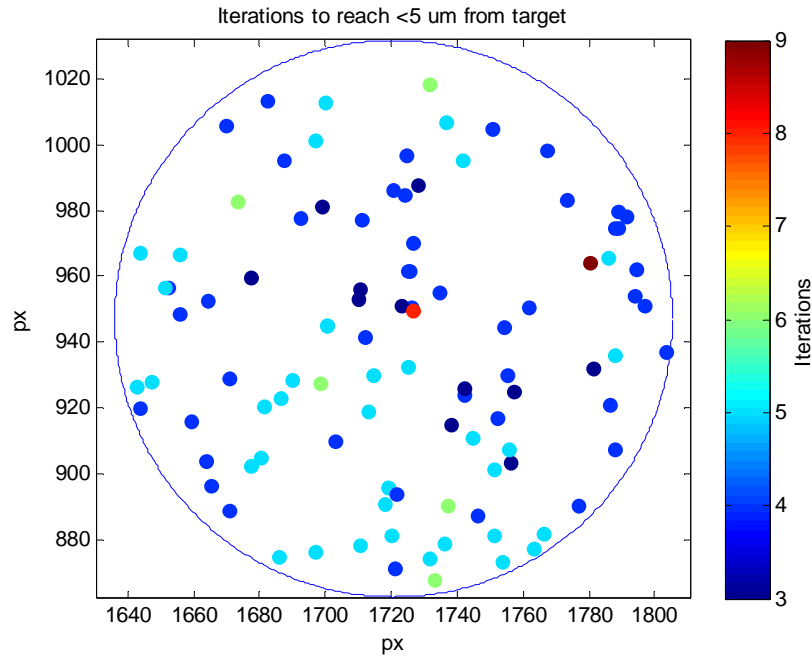

Figure 9 - Target Convergence test results within the patrol region

The convergence to a target can be seen visually be plotting distance remaining vs iteration as seen on the left of Figure 10. The cumulative convergence of all 100 targets can be seen on the right in Figure 10. This cumulative convergence plot is the easiest from which to tell if the metric of 95% convergence to within 5µm in 7 iterations or less is being met. This metric is not a requirement for the project, but serves as a good indicator of system performance. In the example shown, that metric is met within 6 iterations so this positioner performed acceptably.

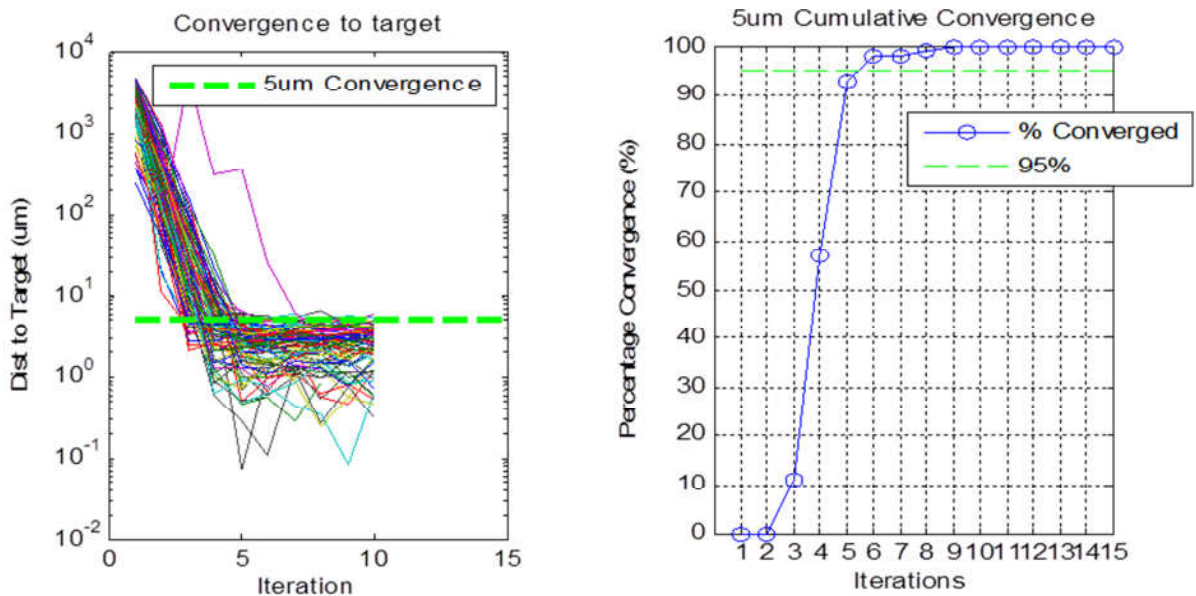

Figure 10 - Distance to target (left) and cumulative convergence (right) as a function of iteration

## 6.3. Other methods of evaluating motor performance

In addition to providing a measure of steps-to-convergence for the motors, the target convergence data can also be used to test and/or determine mathematical models for positioner errors. The simplest model for Cobra motion assumes that each step moves the shaft through a mean angle with a random error: $\psi_{step} = <\psi> + d\psi$. If the movement from each step is independent and the error is Gaussian then the errors follow a random walk. The width of the move error distribution will scale with the square root of the magnitude of the move request:

$$\sigma_{error} = \alpha \sqrt{|\psi_{req}|}, \tag{2}$$

where $\alpha$ is a constant of proportionality, and can be interpreted as the value of $\sigma_{err}$ when $\psi_{req} = 1$ (in whatever units are being used). To model the system, requested moves of $\psi_{req}$ have a resulting error that is drawn from a probability distribution with a characteristic width of $\sigma_{err}$. In practice the error distribution does not always track a perfect random walk model, so a generalized diffusion equation is used to model the error vs. move size:

$$\sigma_{error} = \alpha |\psi_{req}|^{\beta}, \tag{3}$$

For a true random walk, $\beta = 0.5$. When $\beta > 0.5$, the error is "super-diffusive" and when $\beta < 0.5$, it is "sub-diffusive." Both sub- and super-diffusive errors were observed with some of the motors, but a satisfactory physical explanation for super- or sub-diffusive error distributions has yet to be determined.

The error diffusion model for motor movement assumes that the error distribution is symmetric about zero (in other words, the resulting movements are symmetric about the requested move). In early EM1 testing, it was observed that a plot of error vs request angle (see Fig. 11, top panel) for a given target convergence data set would follow a linear trend through the origin. This was the result of a multiplicative error in the motor velocity. The slope from a linear fit (constrained to pass through the origin) is used to correct the velocity calibration for the next target convergence test. The residuals from this fit (Fig. 11, bottom panel) can be used as the errors used to calculate $\alpha$ and $\beta$.

The validity of the model can be tested by normalizing the angular errors, $\Delta\psi$, with $1/|\psi_{req}|^{\beta}$ and verifying that the distribution width of $\Delta\psi|\psi_{req}|^{\beta}$ is independent of request angle. Qualitatively, the trend can be seen in Figure 11 (bottom) at angles below 1 radian, where the envelope of the errors grows with increasing request angle. At request angles greater than 1 radian all of the data points are first moves from the origin, and that the residual errors have some dependence on request angle. For first moves, the mean residual error is the integrated error of the velocity calibration, so the slope is a direct measure of the velocity calibration errors at the position of the motor at the end of the first move.

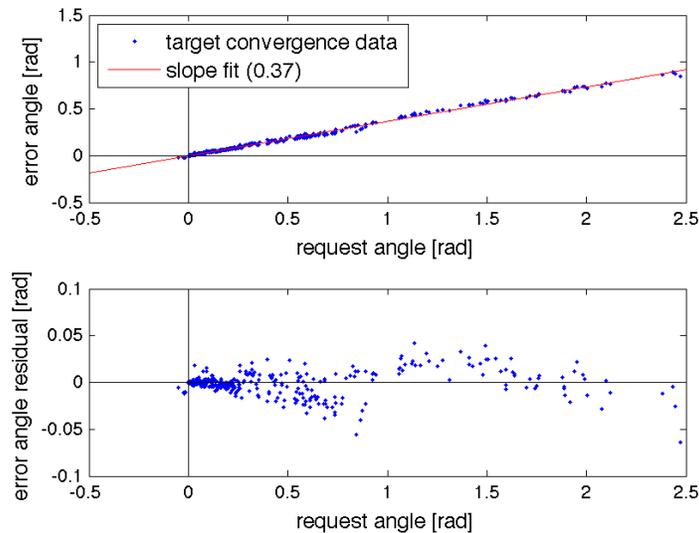

Figure 11 - Example move error vs request angle

Figure 12 shows the (α, β) values determined from target convergence testing of five EM1 motors over several separate tests. In each case, the damping correction was applied to the target convergence data, but there was no attempt to correct for position-dependent velocity calibration errors. Errors in velocity calibration will affect determination of (α, β). Over the course of many tests, under a variety of conditions, the diffusion model parameters (α, β) are sufficiently repeatable to identify positioners when both Stage 1 and Stage 2 parameters are known.

The error diffusion model for Cobra motors allows ranking of positioners in a four parameter space ($\alpha_1, \beta_1, \alpha_2, \beta_2$). In general, smaller values of α are always favorable. For larger values of α, larger values of β are favorable because the "unit move" of 1 radian is larger than the typical move request angle.

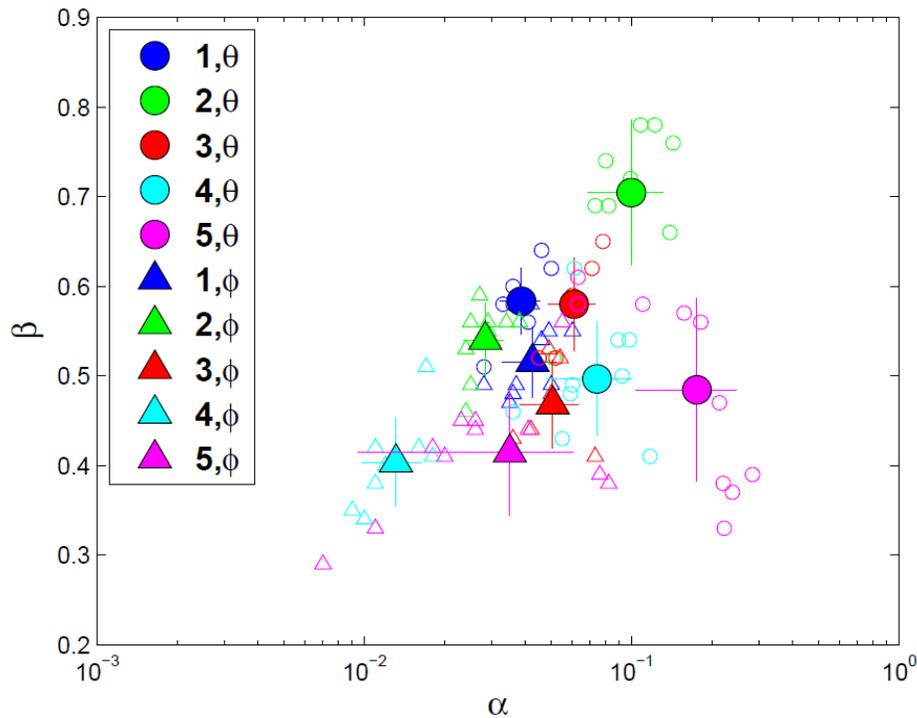
Figure 12 - Alpha-beta results for a sample of EM1 motors

### 6.4. Tilt and focus error testing

The EM test stand constructed at Caltech has the capability to measure fiber centroids, fiber tilt errors and fiber focus errors for an entire module. For centroiding, the concept is to use a compact test stand, which is desirable for ease of use. Imaging the fibers is an important function of the test stand, but it cannot be achieved using the optics that are anticipated on the telescope, which image the PFI plane from about 15 meters away. Instead, the fibers will be back-illuminated with their full numerical aperture (NA) =0.22, which is then increased by the microlenses to NA = 0.25. This large NA enables an imaging lens to be placed within 1.2m of the fiber plane. This stand is shown in Figure 13. A back-illuminated, high bit-count CCD is thought to be needed to achieve the desired precision of 1µm. An Andor 936 camera was chosen, which requires an f=135mm lens to image the fiber plane at the designed distance. Achieving sufficiently large PSFs requires an f/45 lens; a Schnieder Compo non-S lens was chosen.

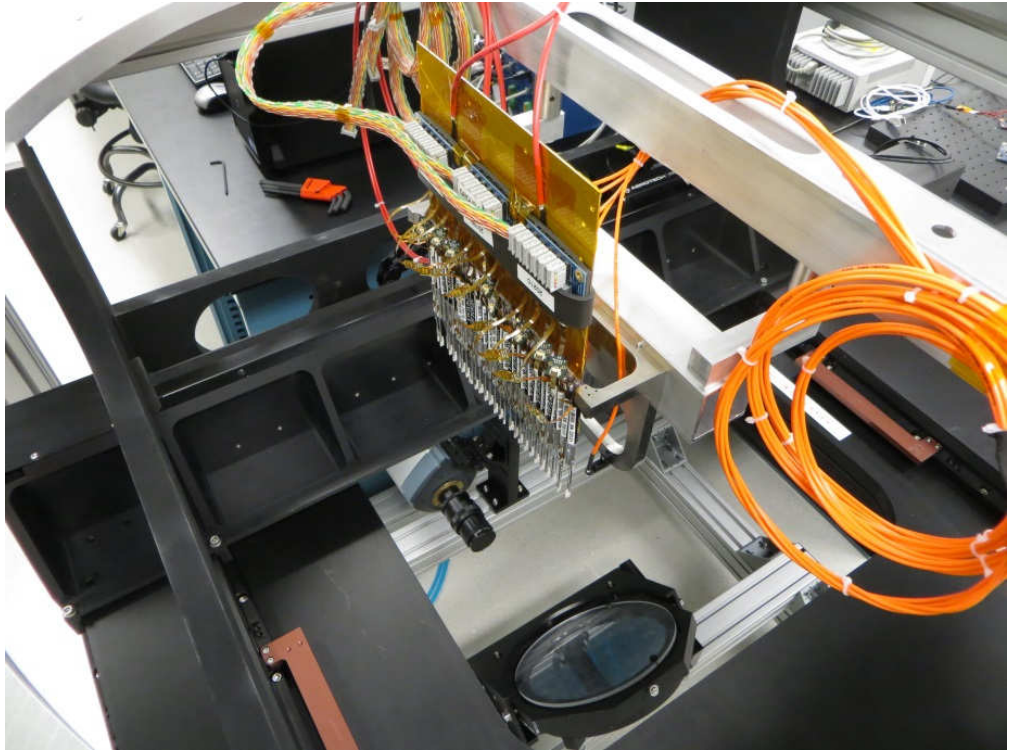
Figure 13 - Caltech Test Stand with EM module installed

An all-optical measurement is required for both focus and angle. This constraint is necessary on angle because the requirement is on the optical axis of the propagated light. This constraint is necessary for focus because the end of the fiber is behind the microlens and mechanically inaccessible.

This concept envisioned optical systems, called "gauges", mounted on an x/y stage; these optical systems are designed to measure focus and angle of the Cobras. Figure 14 shows the concept for angle gauge. The Cobra fiber is placed in the focal plane of a lens; and a camera is placed in the pupil plane, which is in the other focal plane. In this arrangement, ray angle in the fiber plane is converted to ray position in the camera plane. Furthermore, this conversion is insensitive to lateral position and focus in the fiber plane.

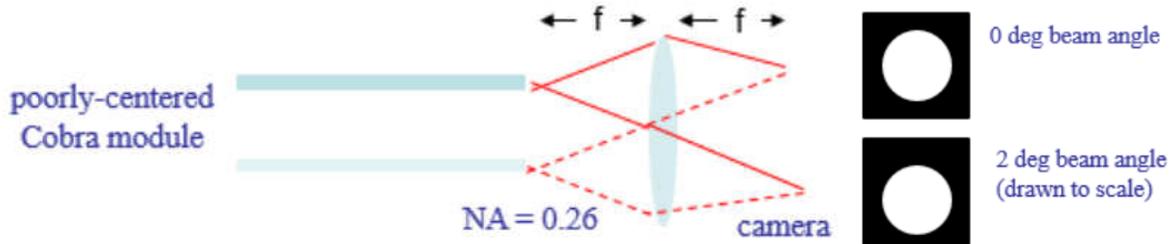
Figure 14 - Conceptual design of the angle gauge

Figure 15 shows the concept for the focus gauge, which is a 2x2 Shack-Hartmann sensor. The Cobra fiber is placed in the focal plane of a lens, and a 2x2 array of lenses is placed in the other focal plane. A camera is then placed in the focal plane of the lens arrays. The 2x2 array of lenses produce a 2x2 array of spots; the separation of the spots measures the focal position of the Cobra fiber. This arrangement is intended to be insensitive to lateral displacement and defocus of the Cobra fiber.

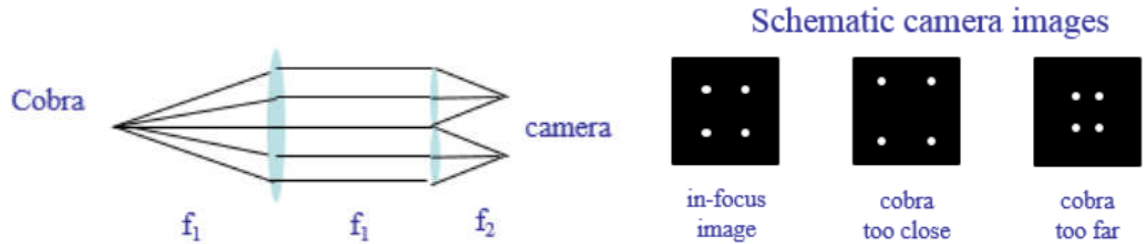

Figure 15 - Conceptual design of the focus gauge

To test an entire module the angle and focus gauges were mounted onto an x/y stage that could rapidly scan across the field of Cobras measuring angles and focus one-by-one. Calibration testing of the gauges showed they have an accuracy of 0.14° for angle and 2.8μm for focus. These are sufficient for characterizing the expected errors in the EM Cobras.

## 7. NEXT STEPS

More testing of the EM Cobras is still required before the start of the production build of PFS. To date only 7 of the 28 Cobras have been evaluated for target convergence. Future tests will also include JPL built motor drive electronics that will meet the packaging constraints of PFS as well as run all 28 Cobras simultaneously from a single board. The tilt and focus error evaluation of Cobras described in section 6.4 will also be completed prior to the start of the production phase.

## 8. SUMMARY

In preparation for the full-scale production of 2550 Cobra fiber positioners an Engineering Model (EM) was built to evaluate the reliability and performance of the positioner design. The EM build was also used to validate the assembly and testing techniques that will be necessary to assemble PFS in an efficient manner. Many design and process issues were discovered and corrected during the development of the EM Cobras. Preliminary evaluation of the performance of the Cobra fiber positioners has yielded positive results. At the successful conclusion of the test campaign the full-scale production of 2550 Cobra fiber positioners will commence.

## 9. ACKNOWLEDGEMENTS

Part of this research was carried out at the Jet Propulsion Laboratory, California Institute of Technology, under a contract with the National Aeronautics and Space Administration.

Funding support was also provided by the Kavli Institute for the Physics and Mathematics of the Universe (WPI), The University of Tokyo.

## REFERENCES

[1] Sugai, H., et al, "Prime Focus Spectrograph – Subaru's Future", Proc. SPIE 8446, Ground-based and Airborne Instrumentation for Astronomy IV (2012)
[2] Fisher, C., Braun, D., Kaluzny, J., Haran, T., "Cobra – a Two-Degree of Freedom Fiber Optic Positioning Mechanism," IEEE Aerospace Conference paper #1185, Version 5 (2009)
[3] http://www.newscaletech.com
[4] Fisher, C., et al, "Developments in high-density Cobra fiber positioners for the Subaru Telescope's Prime Focus Spectrometer", Proc. SPIE 8450, Modern Technologies in Space- and Ground-based Telescopes and Instrumentation II, 845017 (2012)
[5] B. K. P. Horn, "Closed-form solution of absolute orientation using unit quaternions," J. Opt. Soc. Am. A 4,629-642 (1987)